# A transient radio source consistent with a merger-triggered core collapse supernova

D. Z. Dong[1*], G. Hallinan[1], E. Nakar[2], A. Y. Q. Ho[1,3,4], A. K. Hughes[5], K. Hotokezaka[6], S. T. Myers[7], K. De[1], K.P. Mooley[1,7], V. Ravi[1], A. Horesh[8], M.M. Kasliwal[1], S.R. Kulkarni[1]

**Affiliations:**

[1] Division of Physics, Mathematics, and Astronomy, California Institute of Technology, Pasadena, California 91125, USA.

[2] School of Physics and Astronomy, Tel Aviv University, Tel Aviv, 69978, Israel.

[3] Department of Astronomy, University of California, Berkeley, 94720, USA.

[4] Miller Institute for Basic Research in Science, Berkeley, CA 94720, USA.

[5] Department of Physics, University of Alberta, Edmonton AB T6G 2E1, Canada.

[6.] Research Center for the Early Universe, Graduate School of Science, University of Tokyo, Bunkyo-ku Tokyo 113-033, Japan.

[7] National Radio Astronomy Observatory, Socorro, NM 87801, USA.

[8] Racah Institute of Physics, The Hebrew University of Jerusalem, Jerusalem, 91904, Israel.

*Correspondence to: dillondong@astro.caltech.edu

**Abstract:**

A core-collapse supernova occurs when exothermic fusion ceases in the core of a massive star, typically due to exhaustion of nuclear fuel. Theory predicts that fusion could be interrupted earlier, by merging of the star with a compact binary companion. We report a luminous radio transient, VT J121001+495647, found in the Very Large Array Sky Survey. The radio emission is consistent with supernova ejecta colliding with a dense shell of material, potentially ejected by binary interaction in the centuries prior to explosion. We associate the supernova with an archival X-ray transient, which implies a relativistic jet was launched during the explosion. The combination of an early relativistic jet and late-time dense interaction is consistent with expectations for a merger-driven explosion.

**Main Text:**

Most massive stars (those > 8 solar masses, $M_\odot$) are born in close binaries, within which expansion of one star during its evolution can lead to mass transfer with the companion *(1,2)*. In some systems, the faster evolving (more massive) star explodes as a supernova, leaving behind a compact object (neutron star or black hole) remnant in a close orbit with its companion. When the companion (second star) later expands, it transfers mass in the other direction, onto the compact object. Systems of this type with wide orbits have been observed in the Milky Way *(3)*. Those with closer orbits undergo unstable mass transfer, causing the compact object to spiral into the atmosphere of the massive star, forming a common envelope binary.

During the common envelope phase, the outer atmosphere of the donor star becomes unbound, forming a dense and expanding toroidal shell around the binary *(4)*. The physics of the common envelope are difficult to model. Some inspirals halt before reaching the donor's core. This process is a leading candidate for producing the close double-compact-object binaries detected by gravitational wave observatories *(5)*. Other systems are expected to spiral inwards until the compact object reaches the star's core. Theory predicts that some of these systems tidally disrupt the core, forming a rapidly accreting neutrino-cooled disk *(6)*. This energetic accretion is predicted to launch a jet and cause a merger-driven explosion *(6,7,8)*.

We performed a systematic blind search for radio transients in the Very Large Array Sky Survey (VLASS) *(9)*. We identified and followed up luminous point sources associated with galaxies closer than 200 megaparsec (Mpc) that are present in the first half-epoch (Epoch 1.1; Sep 2017 - Feb 2018) of the survey but absent from the earlier (1994-2005) Faint Images of the Radio Sky at Twenty-Centimeters (FIRST) survey *(10, 11)*. The most luminous source we identified was VLASS transient VT J121001+495647 (hereafter abbreviated VT 1210+4956), located in an off-nuclear region of the dwarf star-forming galaxy SDSS J121001.38+495641.7 (fig. 1) *(12)*. This galaxy has stellar mass ~ $10^{9.2}$ $M_\odot$, specific star formation rate ~0.25 $M_\odot$ $Gyr^{-1}$ $M_\odot^{-1}$ and an abundance of elements heavier than helium (metallicity) of ~80% the solar value *(12)*.

We observed two follow-up epochs with the Karl G. Jansky Very Large Array (VLA), finding a peak radio luminosity of 1.5 x $10^{29}$ erg $s^{-1}$ $Hz^{-1}$ at 5 GHz (fig. 1) *(11)*, ten times more luminous than any other non-nuclear transient found in our search. We obtained an optical spectrum of the radio transient location using the Low Resolution Imaging Spectrometer on the Keck I telescope, which exhibits a hydrogen-alpha (Hα at 6563Å) emission line with luminosity (7.3 ± 0.3) x $10^{38}$ erg $s^{-1}$ and full-width half maximum ~1340 ± 60 km $s^{-1}$ (fig. 2) *(11)*. This emission line implies a massive, ionized outflow associated with the radio source *(13,14)*, and its redshift $z$ = 0.03470±0.00003 confirms its association with the host galaxy, which is at $z$ = 0.03472±0.00003 *(12)*. In addition to the broad component, we observe spectrally unresolved (narrow) emission lines consistent with a co-located star forming region. The environment, broad line, and high radio luminosity indicate a likely association with the explosion of a massive star.

Radio emission from supernovae is powered by synchrotron emission from fast ejecta driving a shock into the ambient circumstellar medium (CSM) *(15)*. The most luminous events require exceptionally fast material, such as relativistic (close to the speed of light) jets driven by central compact-object engines, or interaction with a particularly dense CSM. All known supernovae with radio emission peaking at ≳$10^{29}$ erg $s^{-1}$ $Hz^{-1}$ are high-luminosity examples of transient classes that involve central engines *(e.g. 16—21)*. Supernovae featuring dense CSM interaction are typically over an order of magnitude less luminous. Given the high luminosity of VT 1210+4956, we checked for early-time signatures of a central engine by searching archival optical, X-ray, and gamma-ray transient catalogs for a counterpart source *(11)*. This search yielded one match: GRB 140814A, an unusual soft X-ray burst detected by the Monitor of All Sky X-ray Image (MAXI) instrument on the International Space Station, using its Gas Slit Camera (GSC) *(22)*.

GRB140814A was detected by the GSC at 2-4 keV and 4-10 keV in a 15 ± 3s window centered at 07:12:23 Universal Time on 2014 August 14, with a similar flux in both bands. It was not detected in a simultaneous observation with a similar sensitivity in the GSC 10-20 keV band, suggesting a characteristic energy of ~5 keV. The position of VT 1210+4956 is consistent with the MAXI data and implies a short, fading burst that occurred at the beginning of the GSC's ~40 second transit (fig. 3) *(11)*. We estimate a false alarm probability for the spatial association of (1.2 to 4.8) × $10^{-3}$ *(11)*. We performed several additional consistency checks: (a) the shock properties of VT 1210+4956 are consistent with an explosion on the date of GRB 140814A, (b) upper limits from contemporaneous optical observations do not rule out a stripped envelope supernova at the location of VT 1210+4956 and time of GRB 140814A, and (c) alternative classes of X-ray transients are not consistent with the observational data *(11)*.

Our association of VT 1210+4956 with GRB 140814A implies that the X-ray emission had a peak 2-10 keV luminosity of ~4 × $10^{46}$ erg/s. This combination of high luminosity, short duration, and soft spectrum is unlike other X-ray transients with measured luminosities. Known transients with similar spectral peaks and durations, such as the shock breakout of supernova SN 2008D *(23)*, are more than 3 orders of magnitude less luminous. Transients with similar luminosities, such as X-ray flashes and low-luminosity GRBs, typically maintain this luminosity for at least an order of magnitude longer duration *(24)*, or peak at $\gtrsim$ 1 order of magnitude higher energy *(25)*. The luminosity and duration of MAXI 140814A implies a relativistic outflow with Lorentz factor $\gamma \gtrsim$ 2.5, and the characteristic energy of the photons implies a jetted geometry *(11)*. Producing such a relativistic jet requires the presence of a central engine at the time of explosion.

Unlike the early relativistic jet, the radio emission detected in our follow-up epochs (observed >3 years after GRB140814A) is powered by a low velocity shock propagating in a dense, extended CSM. The spectral peaks of the radio spectra are due to synchrotron self-absorption rather than free-free absorption, suggesting that the CSM is asymmetric, observed along a lower-density line of sight *(11)*. The peak luminosities and frequencies imply forward shock radii $R \sim 9 \times 10^{16}$ cm, post-shock magnetic fields $B \sim 0.35$ G, and energies dissipated in the shock $U \sim 7 \times 10^{49}$ erg *(11)*. The change in $R$ between follow-up epochs implies a forward shock velocity of $\sim 1800$ km/s *(11)*. This velocity, which is similar to the width of the broad H$\alpha$ line, implies a high density of $\sim 10^{6}$ cm$^{-3}$ for the CSM swept-up in the ~1 year between follow-up epochs. This density is sufficient to allow the shocked gas to cool, on a timescale of ~1 year, from a shock-heated temperature of $\sim 10^{7}$ K to a $\sim 10^{4}$ K dense shell, cool enough to produce the H$\alpha$ emission *(26)*. Compared to known explosive radio transients, VT 1210+4956 has a highly energetic shock propagating at low velocity, and a high CSM density at large radius (fig. 4).

Measurements of the CSM density as a function of radius trace the rate and timing of pre-explosion mass loss. Accounting for both the cool H$\alpha$ emitting gas and the hot synchrotron emitting gas, we find a lower limit to the total swept-up CSM mass of $\gtrsim$1 M$_\odot$ *(11)*. The high density and total mass at a radius of ~$10^{17}$ cm requires a pre-explosion mass loss rate $\dot{M} \approx 4 \times 10^{-2}$ ($v$/100 km s$^{-1}$) M$_\odot$ yr$^{-1}$, where $v$ is the pre-shock CSM velocity *(11)*. This is over an order of magnitude higher than the densest observed stellar winds and requires a pre-supernova eruption

*(27)*. With these observed and inferred constraints on the CSM velocity, we find that the eruption occurred a few centuries before the explosion (fig. 4) *(11)*.

Similar pre-explosion eruptive mass loss has been inferred from the dense and extended CSM around a few peculiar supernovae, including SN 2014C *(28 — 30)* and SN 2001em *(31,32)*, which both transitioned from stripped envelope (type Ib/c) to interacting (type IIn) spectral classifications. Several models have been proposed to explain the synchronization of this eruptive mass loss with core collapse, including nuclear burning instabilities *(32,33)*, binary interaction timed coincidentally with core collapse *(28,30)*, and merger driven explosions *(6,7,28)*. For VT 1210+4956, the detection of a central engine allows us to distinguish between these scenarios, and the order of magnitude higher radio luminosity and shock energy suggests the possibility of a different origin to SN 2014C and SN 2001em.

Nuclear burning instabilities strong enough to produce VT 1210+4956's pre-explosion mass loss rate are predicted to occur in only the final few years before core collapse *(33)*, so would produce shells that are over an order of magnitude more compact at the time of interaction than we observed. Binary mass transfer is common in massive stars and is more consistent with the early eruptive mass loss. Roughly 70% of massive stars are found in orbits that will eventually lead to mass transfer, with an estimated 1/3 of these interactions leading to common envelope inspiral *(1,2)*. Such interactions are predicted to drive mass loss at or above the rate inferred for VT 1210+4956 *(27)*. Unlike in single star models, the mass is ejected in the plane of the binary *(4, 7)*, providing a natural explanation for the inferred asymmetry of the CSM.

Binary interactions can occur at any time during the life of a star and can thus produce shells at any radius. However, the delay time between eruption and core collapse constrains the specific type of interaction. At the near solar metallicity of VT 1210+4956's host galaxy, most interactions are expected to occur while the mass donor is undergoing fusion of hydrogen or helium, roughly $10^4$ - $10^7$ years before core collapse *(34)*. Interactions synchronized coincidentally within ~$10^2$ years of core collapse are expected to be extremely rare *(34)*, though uncertainties in the rates from binary population synthesis modeling allow this to be a viable scenario for previous events such as SN 2014C *(28, 30)*.

The synchronization is more naturally explained if the interaction itself triggers the core collapse. To do so, an inspiraling object must disrupt the donor star's core. A non-compact merging body (e.g., a main sequence star) is unable to do so, as it would replenish fuel in the core, producing a rejuvenated massive star *(35)*. In contrast, an inspiraling neutron star or black hole is capable of tidally disrupting the core, leading to a merger driven explosion *(6,7)*. During the inspiral phase, the compact object is expected to eject mass from the donor star at a similar rate to our calculated value *(36)*. When it reaches the core, theory predicts the formation of an accretion disk and launching of a jet *(6,7)*. Of the models we consider for eruptive mass loss, this scenario is most consistent with the jetted X-ray transient.

The high mass loss rate we infer occurred centuries before explosion, much longer than the dynamical timescale of the inspiraling compact object, which may indicate that the donor star had an envelope with a steep density profile. This would be theoretically expected for a star undergoing core helium fusion, for example *(36)*. A steep density profile is thought to be required for merger during the dynamical inspiral phase *(36, 37)*. Flatter density profiles would lead to full envelope ejection before merger, producing close compact object – evolved star binaries. These binaries may subsequently evolve into double compact object systems, with orbits close enough to merge within the lifetime of the Universe, and therefore contribute to gravitational wave events *(5, 36, 37)*. Our proposed scenario for VT 1210+4956 is an alternative outcome to the formation of such systems.

On the basis of our blind search, we estimate a rate of (1 to 8) x $10^{-8}$ $Mpc^{-3}$ $yr^{1}$ for transients with similar 3GHz luminosities to VT 1210+4956 *(11)*. However, given the continuous distributions of stellar binary periods and mass ratios *(1, 2)*, there may be a wider range of delays between interaction and core collapse, thus producing CSM shells at smaller or larger radii. If this is the case, then these events may be more easily identified at other frequencies, so we regard this rate as a lower limit on the rate of merger-driven explosions.

**Acknowledgments:** We thank the NRAO staff who made the VLASS possible and the VLASS Survey Science Group. We thank J. Fuller and J. Lux for useful discussions, and the reviewers for helpful comments. Our results are based on data from the Karl G. Jansky Very Large Array, which is operated by the National Radio Astronomy Observatory (NRAO). The NRAO is a facility of the National Science Foundation operated under cooperative agreement by Associated Universities, Inc. This research has made use of MAXI data provided by RIKEN, JAXA, and the MAXI team. Some of the data were obtained at the W. M. Keck Observatory which is operated as a scientific partnership among the California Institute of Technology, the University of California, and the National Aeronautics and Space Administration. Keck Observatory was made possible by the generous financial support of the W. M. Keck Foundation. We recognize and acknowledge the cultural role and reverence that the summit of Maunakea has always had within the indigenous Hawaiian community. We are most fortunate to have the opportunity to conduct observations from this mountain. **Funding:** DZD and GH are supported by NSF grant AST-1654815. GH, DZD, and AH. acknowledge the United States – Israel Binational Science Foundation grant 2018154. AH acknowledges support by the I-Core Program of the Planning and Budgeting Committee and the Israel Science Foundation, and support by ISF grant 647/18. AYQH and KD were supported by the GROWTH project funded by the National Science Foundation under PIRE Grant No. 1545949. AKH is supported by NSERC Discovery Grants RGPIN-2016-06569 and RGPIN-2021-04001. AYQH is supported by the Miller Institute for Basic Research in Science at the University of California Berkeley. KH is supported by the JSPS Early-Career Scientists Grant No. 20K14513. SM was supported by the NRAO. SRK acknowledges support from the Heising-Simons Foundation. **Author contributions:** GH, DZD, and KPM designed the transient search strategy. DZD implemented the transient search pipeline, with help from AKH in visual vetting and AYQH in matching the MAXI transient. DZD, GH, and STM obtained the radio and optical follow-up observations with the help of telescope staff at the VLA and Keck. DZD processed and analyzed the radio data with contributions from STM. DZD processed and analyzed the optical data and estimated the transient rates. EN led the analysis of the X-ray data with help from DZD. DZD and GH wrote the paper with contributions from EN, KH and all coauthors. **Competing interests:** We declare no competing interests. **Data and materials availability:** VLASS quicklook images are available at https://archive-new.nrao.edu/vlass/quicklook/ . The VLA follow-up observations are available from the VLA archive https://archive.nrao.edu/ under project code 19A-422. The FIRST data were taken from https://third.ucllnl.org/cgi-bin/firstcutout. The Keck observations are available at https://koa.ipac.caltech.edu/ with observation ID LB.20180413. The Hubble Space Telescope image was taken from the archive https://hla.stsci.edu/ under proposal ID 13493. Our analysis and model fitting code is available at https://github.com/Dillon-Z-Dong/VT1210 and archived on Zenodo *(39)*.

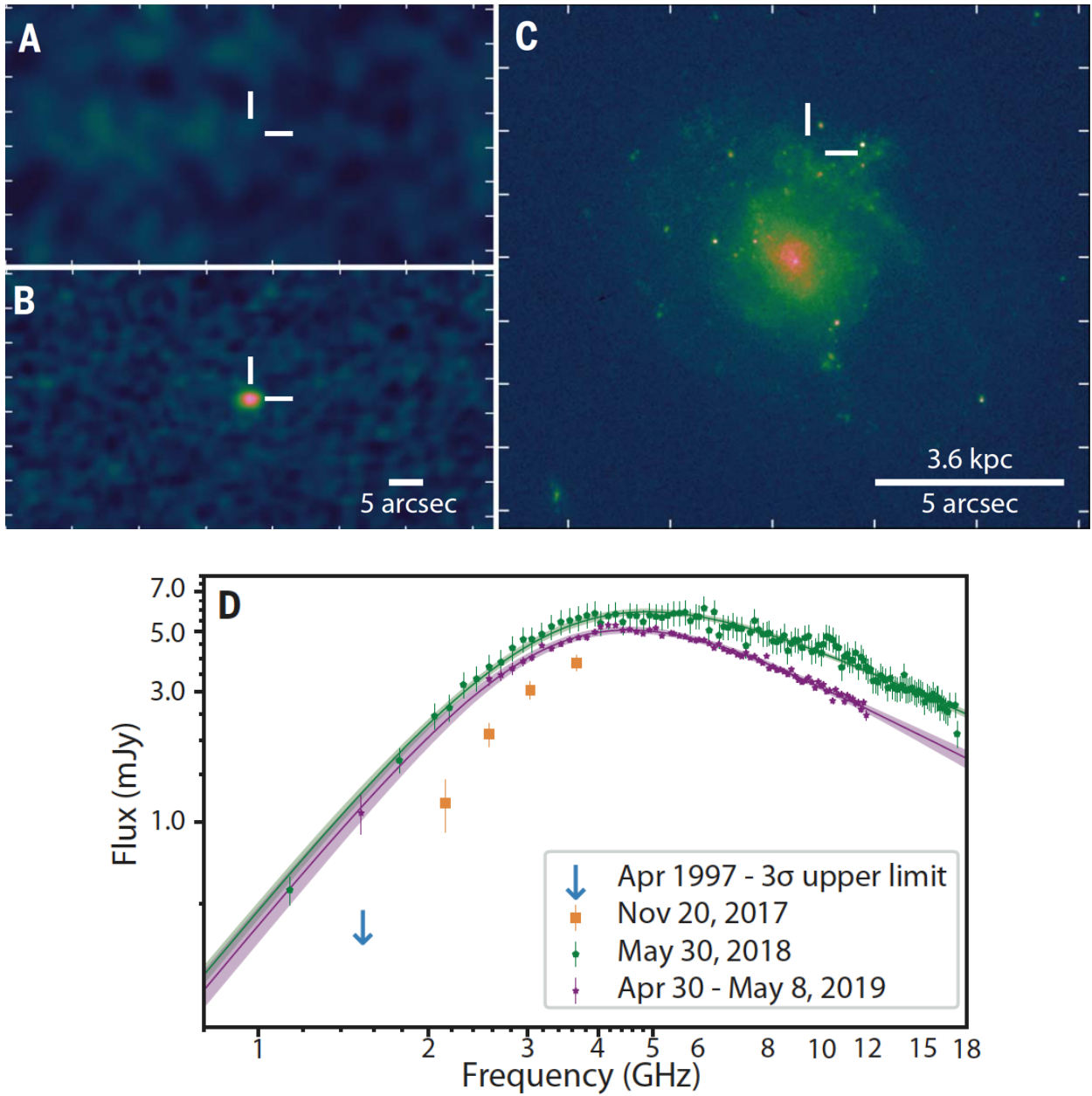

**Fig 1.** The luminous radio transient VT 1210+4956. **A.** Non-detection in the FIRST survey at 1.4 GHz, with a 3σ upper limit of 0.41millijansky (mJy) on 1997 April 17. **B.** Detection in VLASS at 3 GHz at 2.7 ± 0.1 (stat) ± 0.5 (sys) mJy on 2017 November 20, 20.59 years after the FIRST observation, at right ascension 12h10m01.32s, declination +49°56'47.006" (indicated by white crosshairs in panels A-C). **C.** Optical image of the location of VT 1210+4956 taken from the Hubble Space Telescope archive (PI: T. Schrabback). **D.** The radio spectrum of VT 1210+4956 measured from the VLASS observation and follow-up epochs observed with the VLA, plotted with 1σ uncertainties. Follow-up epochs are fitted with a synchrotron self-absorption model *(11)*.

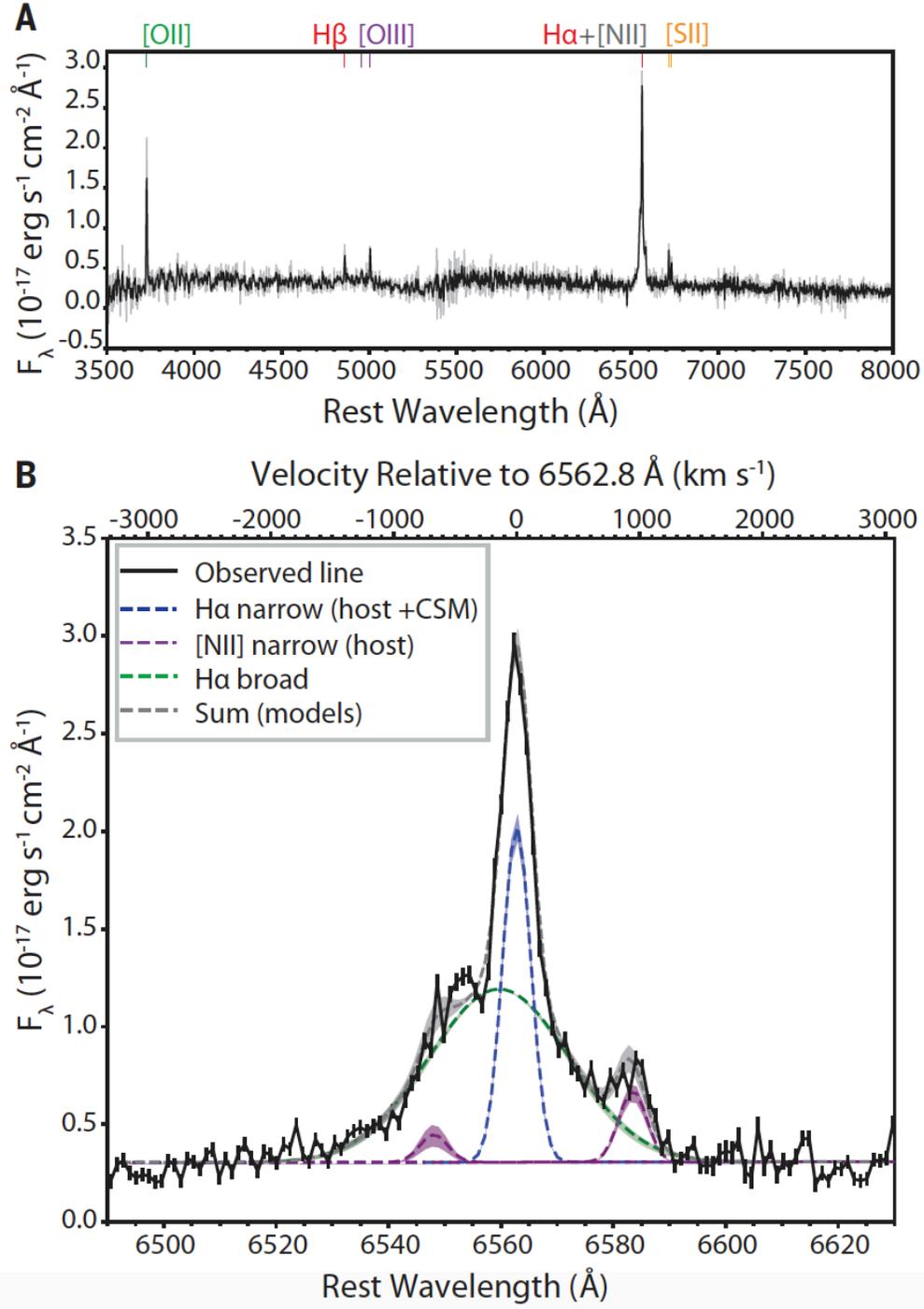

**Fig 2. Optical spectrum of VT 1210+4956.** Taken with Keck/LRIS on 2018 April 13, ~4.45 months after the detection in VLASS. **A.** The full spectrum, including spectrally unresolved emission lines from the host galaxy. Unbinned data are shown in gray; data shown in black were smoothed with a 3Å boxcar kernel. **B.** Details of the Hα & [N II] part of the spectrum, with Gaussian models fitted to the lines. The narrow Hα and [N II] lines are unresolved, while the broad component has a FWHM of 1340 ± 60 km/s. The best fitting model parameters and their uncertainties are listed in Table S1.

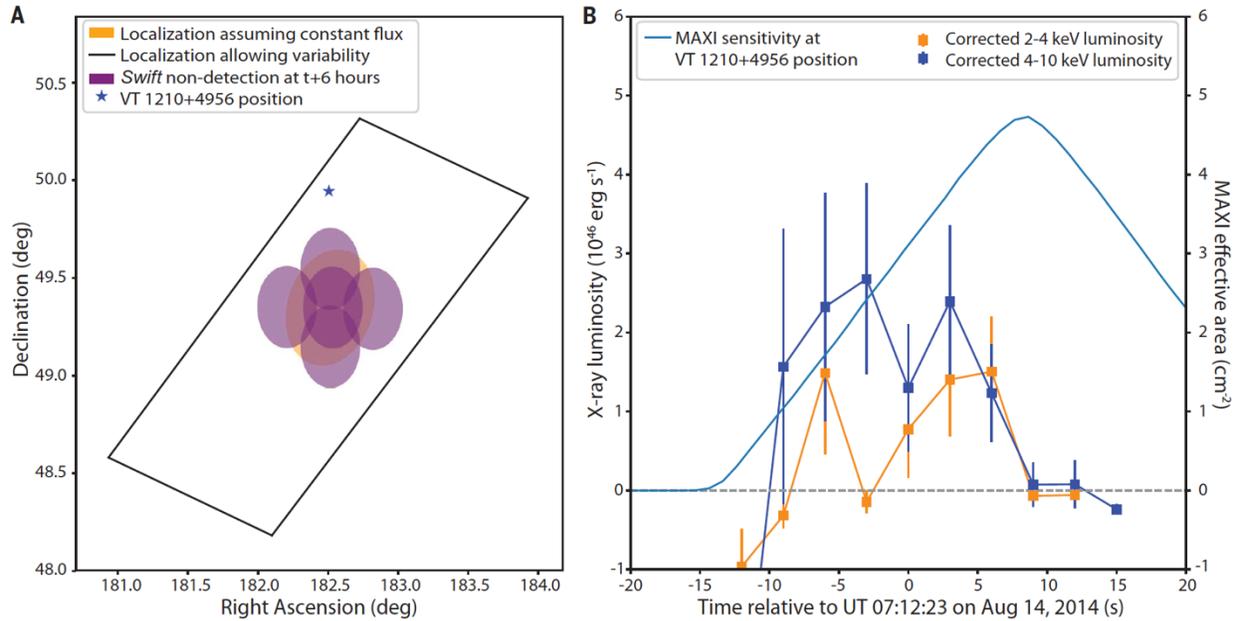

**Fig 3. The X-ray transient MAXI 140814A which we associate with VT 1210+4956** (observed 3.268 years before VLASS detection). **A.** The location of VT 1210+4956 compared to published MAXI localizations: orange ellipse assuming constant flux and black rhombus allowing for variability. This position is consistent within 1σ of the expectation from the lightcurve in panel B *(11)*. The purple ellipses show fields with non-detections in Swift observations, with upper limits of 60,000 times fainter than the MAXI 140814A ~6 hours after the burst *(11)*. **B.** MAXI 2-4 keV (orange) and 4-10 keV (blue) light curves, corrected for instrumental sensitivity at the location of VT 1210+4956 *(11)*.

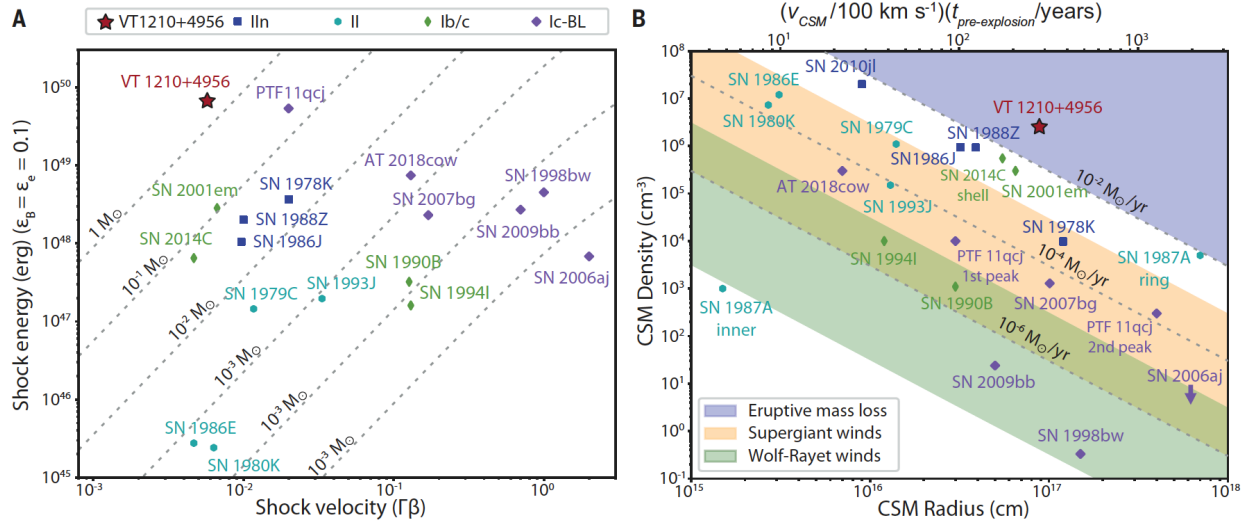

**Fig 4. Comparison of inferred shock properties for VT 1210+4956 and other luminous radio supernovae** (sources listed in Table S5). **A.** Shock energies derived from broadband radio spectra vs shock velocities. The dashed curves show constant shocked mass, assuming that the shock energy is equal to the kinetic energy. **B.** CSM density vs CSM radius with values derived from optical and radio spectra. The upper axis gives the time between mass loss and explosion, normalized to a CSM velocity of 100 km/s. The dotted and dashed lines indicate constant mass flow rates $\dot{M}/v$, with the colored shaded regions indicating approximate ranges of values for winds and eruptive mass loss from different types of systems *(27)*.

**Materials and Methods**

Transient Identification in the VLA Sky Survey

We assembled a catalog of sources detected with a significance of ≥ 5σ in the quick-look images for the first half-epoch of VLASS (Epoch 1.1) *(9)* using the source extractor PYBDSF *(39)*. Within this catalog, we flagged the set of VLASS source separated by >10" from the nearest catalogued source in FIRST *(10)* as our initial set of transient candidates. Through visual inspection of all candidates, we selected those that are unresolved and have no identifiable uncatalogued counterpart in FIRST. To identify the population associated with nearby galaxies and exclude background sources (e.g., flaring active galactic nuclei (AGN)), we use the fact that background sources are isotropically distributed *(40)*, while known classes of extragalactic transients are concentrated within a few exponential radii ($r_{exp}$) of their host galaxies *(41)*. In our sample, we find an excess of transient candidates within $2r_{exp}$ of their host galaxies, where the $r_{exp}$ of a galaxy is taken to be the median among all detected bands in Data Release 15 of the Sloan Digital Sky Survey (SDSS) *(12)*. There are 20 sources within $2r_{exp}$ of a d < 200 Mpc galaxy, with an expected isotropic background rate of 0.6, given the full set of visually confirmed VLASS sources not in FIRST.

Of these 20 sources, 12 have astrometry consistent with their host galaxy's nuclei, while 8 are in off-nuclear positions. To avoid likely contamination from AGN, we focused on the off-nuclear transients. Follow-up optical spectra of the ~1 arcsecond regions surrounding each radio transient revealed that 4 of the 8, including VT 1210+4956, have ~1000 km/s hydrogen emission lines at the redshift of the host galaxy, indicating that the transient radio sources are associated with their apparent host galaxies. These lines are too broad to be due to star formation *(42)*. Among these 4 sources, VT 1210+4956 was the most radio luminous by an order of magnitude (with a peak radio luminosity of ~2 x $10^{29}$ erg/s/Hz), and the only one lacking other broad emission lines in its optical spectrum.

VT 1210+4956 is located at right ascension 12h10m01.3s, declination +49d56'47.0'' in a star forming region on the outskirts of the galaxy SDSS J121001.38+495641.7. This position is derived from VLA follow-up imaging described below, and has an uncertainty smaller than 0.1''. From broadband spectral energy distribution modeling listed in the SDSS data products *(12)*, this galaxy has a stellar mass of ~$10^{9.2}$ $M_\odot$, a star formation rate of ~0.3 $M_\odot$ yr$^{-1}$, and a metallicity of 0.8 $Z_\odot$.

Optical follow-up spectroscopy

Five to 20 months after the VLASS observations and ~10-20 years after non-detections in FIRST, we obtained optical spectra at the location of the 8 off-nuclear radio transients with the Low Resolution Imaging Spectrometer (LRIS) on the 10m Keck I telescope on Maunakea. We used the 560nm dichroic to split the light between the red and blue arms. We dispersed the light with the 400/8500 grating in the red arm, and the 400/3400 grating on the blue arm, resulting in a ~7Å FWHM spectral resolution on both arms. To acquire the target locations, we centroided on nearby bright stars and applied an offset corresponding to the difference between the SDSS coordinate of the star and the best-fitting coordinate of the radio transient. We observed each target in spectroscopic mode for 20 minutes, with 2×1 (spatial × spectral) CCD binning in the red arm, and 2×2 CCD binning in the blue arm. These settings correspond to a 3$\sigma$ sensitivity of ∼23 mag (AB) throughout the red arm and a wavelength dependent 3$\sigma$ sensitivity peaking at ∼24 mag (AB) at 5000Å in the blue arm and dropping on either side to ∼23 mag at 4000 Å and the 5600Å dichroic. We reduced the LRIS observations using the data reduction pipeline LPIPE *(43)*.

Four of the resulting spectra are consistent with an underlying star forming region only, while the other four contain additional broad hydrogen and oxygen features at the same redshift. We focus on the spectrum of VT 1210+4956, taken on 2018 April 13, which contains a broad H$\alpha$ line that is blended with spectrally unresolved H$\alpha$ and [N II] lines. We fitted this complex with a model of four Gaussian components, corresponding to the broad and narrow H$\alpha$ lines, and the narrow [N II] host galaxy lines. The narrow line components have central wavelengths fixed to their respective wavelengths at the host galaxy redshift $z = 0.03472$ *(12)*. Their widths were fixed to the 7Å FWHM instrumental resolution, while their amplitudes were allowed to vary. The broad profile was allowed to vary in its amplitude, width (up to 10,000 km/s) and central wavelength (up to 6Å). We additionally marginalized over three nuisance parameters corresponding to i) an overall zero-point offset for the spectrum, ii) an overall shift in the wavelength solution (corresponding to a redshift uncertainty of < 0.002), and iii) a local slope for the continuum emission. We fitted this model using the Markov Chain Monte Carlo code EMCEE *(38, 44)* and find that the data are sufficient to constrain each parameter to a roughly Gaussian posterior probability distribution. The model spectrum is shown in fig. 2, and the best-fitting values are listed in table S1.

Radio follow-up observations

Using the VLA, we took two broadband radio follow-up observations of VT 1210+4956. The first epoch was observed on 2018 May 30 and spans the frequency range 1–18 GHz in five bands with central wavelengths 1.5, 3, 6, 10, and 15 GHz. The second epoch spans 1–12 GHz, and was observed in two scheduling blocks, with the 1.5 and 3 GHz bands observed on 2019 April 30, and the 6 and 10 GHz bands observed on 2019 May 8. We used standard procedures in the radio data reduction package CASA *(45)* to calibrate and image the data. The photometric measurements based on the images in each band are listed in table S2. Where permitted by signal to noise, we increased the frequency resolution by producing sub-band images from groups of independently calibrated spectral windows. After verifying in each sub-band image that the target is detected, unresolved, and not substantially affected by image artifacts, we measured the peak flux using the CASA task *imstat* and estimated the uncertainty by taking the root mean square pixel value in an area of the image with no substantial emission. In both epochs, we detected the peak of the radio spectrum, allowing for standard synchrotron blast wave modeling described below. In our first follow-up epoch, taken ~20.6 years after nondetection at 1.5 GHz in FIRST, we measured a 1.5 GHz source at a flux of 1.22 ± 0.04 mJy (Table S2). After adding the 0.13 mJy RMS noise in the FIRST image in quadrature with our measurement uncertainty, this corresponds to a 9$\sigma$ detection of variability relative to the nondetection in FIRST, confirming that VT 1210+4956 is a transient source.

Associated X-ray burst

We searched the MAXI Gamma Ray Burst (GRB) catalog *(21)*, the INTernational Gamma-Ray Astrophysics Laboratory (INTEGRAL) GRB catalog *(46)*, the Swift GRB catalog *(47)*, and the Open Supernova Catalog *(48)* for counterparts to VT 1210+4956. We considered an archival transient to be a match to VT 1210+4956 if it fell within the astrometric uncertainty of that source's localization region. For MAXI, INTEGRAL and Swift, we used the published uncertainties and localization regions, and for optical supernovae in the Open Supernova Catalog, we took this region to be a circle of radius 2 arcseconds. We excluded the Fermi GRB catalog *(49)*, because the large localization regions and the ~2000 bursts detected before the VLASS detection in 2017 lead to a high false alarm probability (an expected value of ~13 spurious matches per source as estimated by cross matching the catalog with a set of random coordinates). Among the catalogs adopted, we found one match: an association between VT 1210+4956 and the soft X-ray flash GRB 140814A *(21)*.

GRB 140814A was detected by the MAXI Gas Slit Camera (GSC) in the 4-10 keV band during five consecutive 3-second time bins, centered between Universal Time (UT) 07:12:17 and 07:12:29 on 2014 August 14. The integrated 4-10 keV flux in this time window is ~11x greater than the upper limits from the previous transit at UT 5:39 and the subsequent transit at UT 8:43 *(50)*. The burst was simultaneously detected in the GSC's 2-4 keV band in four of the five time bins (fig. 3). There was no detection in the simultaneous observation with the GSC 10-20 keV band, which has similar $1\sigma$ sensitivity to the other two bands *(51)*. This burst is one of 9 MAXI GRBs that has been classified as a MAXI Unidentified Short Soft Transient (MUSST), due to the lack of counterparts in follow-up observations *(52–53)* and the non-detection in the highest energy MAXI band.

Consistency of VT 1210+4956's position with the GRB 140814A localization

Sources detected with the MAXI GSC are localized by the MAXI team by modeling the observed light curve with the position-dependent instrumental response *(21)*. Along the direction of the slit, point sources appear as a roughly Gaussian point spread function (PSF) with a standard deviation of 0.42-0.64 degrees depending on the angle at which they pass over the slit *(51)*. As these sources transit over the slit, their flux is modulated by a triangular effective area function dependent on the orthogonal separation between the source and the slit *(51)*. The observed light curve is proportional to the product of the intrinsic light curve with the effective area. Sources with intrinsically constant flux can thus be localized by fitting a joint model of the triangle function in the time direction and a Gaussian in the direction along the slit *(21)*. For sources that vary during the observation, the light curve cannot be properly represented by the triangle function, which produces a systematically too small localization region. To roughly compensate for this effect, the MAXI team produced a variable source localization region by first fitting a model to the data assuming the flux were constant, then extending the localization in the direction of the scan by a factor of $d\Theta(T_{transit} - T_{duration})/T_{transit}$ where $d\Theta$ is the angular-width of the triangle function, $T_{transit}$ is the time-width of the triangle function, and $T_{duration}$ is the observed width of the transient. This method is estimated to localize the position of GRB afterglows up to a ~0.1 deg systematic error *(21)*.

Figure 3 shows the position of VT 1210+4956 relative to the published MAXI localization regions for constant and variable sources, after incorporating the systematic error. We used the MXSCANCUR function from the MAXI software package *(38, 54, 55)* to compute the angular separation of the VT 1210+4956 position and the center of the constant flux ellipse as a function of time. At the time of the burst the separation is ~0.34 deg, smaller than the $1\sigma$ uncertainty of the Gaussian response *(51)*. We additionally used MXSCANCUR to check that the effective area at the position of VT 1210+4956 is nonzero at the time of the burst detections. We find that this is the case, with all $>1\sigma$ detections occurring during the first half of the sensitivity window (fig. 3).

VT 1210+4956 is located in the variable source region, but not the constant X-ray flux region (fig. 3). This is as expected if the observed duration of the burst (~15s) is shorter than the MAXI transit duration (~40s). To check that this short duration cannot be explained as an artifact of the time-variable GSC sensitivity, we ran a Monte Carlo simulation *(38)*. If the true light curve were constant and longer than the ~40s transit duration, then the observed light curve can be modeled with a top hat function of width >40s, multiplied by the effective area function. We thus fitted the observed light curve with a model product of these functions, allowing the duration and flux of the burst to vary freely and using the effective area function at the center of the constant source localization region. We adopted flat priors in both the duration and flux, with the height constrained to be greater than 0 and less than 5x the highest data point, and the width constrained to be greater than 0 and less than 100s. We fit the 4-10 keV data because it has a higher signal-to-noise ratio than at 2-4 keV. From this simulation, we found that the best fitting top hat model has a duration of $15.1 \pm 2.3$s (where the uncertainty reflects the $50^{th}$-$16^{th}$ and $84^{th}$-$50^{th}$ percentile ranges). Burst durations of >40s are ruled out at the 99% level, consistent with GRB 140814A being located in the variable source region *(38)*.

A location for GRB 140814A outside the constant flux region is reinforced by the non-detection in 0.3-10 keV follow up observations by the Swift X-ray telescope taken ~6 hours after MAXI. These data cover the constant flux localization region (fig. 3) and do not detect any new X-ray sources with an upper limit of $3\times10^{-3}$ counts/s *(53)*. This corresponds to a flux of $\sim10^{-13}$ erg cm$^{-2}$ s$^{-1}$, approximately 60,000x fainter than the MAXI detection in a similar band, a substantially faster fade rate than typical for GRB afterglows at this epoch *(56)*.

False alarm probability

We test the false alarm probability (FAP) for GRB 140814A being located in the variable source region, excluding the constant source region. After incorporating the systematic uncertainty, this region spans an area of ~1.54 deg$^2$. Due to the uncertainty in the intrinsic light curve of the burst and its degeneracies with the localization, we do not attempt to further subdivide this area. We treat all points within this area as being equally likely to be the true position.

VT 1210+4956 was identified in a blind search of the overlap between the VLASS Epoch 1.1 and the FIRST fields, spanning 6195 deg$^2$. The probability that a random point from the searched area falls within the GRB 140814A region is 1.54 deg$^2$ / 6195 deg$^2$ = 2.5 x 10$^{-4}$. Because GRB 140814A is the only MUSST within the Epoch 1.1 / FIRST overlap that occurred before the end of Epoch 1.1 in February 2018, this is the FAP for a cross match between VT 1210+4956 and MUSSTs.

If we relax the softness condition, there are 3 other MAXI bursts (GRB 150110A *(57)*, GRB 120528C *(58)*, and GRB 111024A *(59)*) with no definitive counterparts . The collective area spanned by the localizations of all 4 bursts (including GRB 140814A) is 6.6 deg$^2$, corresponding to a FAP of 1.1 x 10$^{-3}$. Expanding the sample to account for other catalogs of potential matches only changes this result by small amount. The 12 INTEGRAL GRBs *(46)* in the Epoch 1.1/FIRST overlap are localized to a few arcminutes and span a total area of 0.053 deg$^2$. The Swift GRBs *(47)* span 0.6 deg$^2$, and the Open Supernova Catalog supernovae *(48)* span 0.017 deg$^2$ assuming a 2 arcsecond match radius. The joint false alarm probability for all those catalogs combined is 1.2 x 10$^{-3}$. We regard this value as the lower limit of the FAP.

There are 3 other VLASS transients among the sample that are associated with galaxies and have broad Hα lines in follow-up spectroscopy. These transients are ~an order of magnitude less radio luminous than VT 1210+4956. Their optical spectra all contain broad oxygen lines, which VT 1210+4956's spectrum lacks. None of the other 3 are associated with observed X-ray bursts. Their radio and optical properties are also consistent with supernovae interacting with massive CSM shells ejected in the centuries before core collapse. If we consider all four to be part of the same sample, the FAP rises by a factor of 4 to 4.8 x 10$^{-3}$. We consider this to be an upper limit on the FAP.

Peak flux and frequency

To infer properties of the shock from our broadband radio SEDs, we used EMCEE *(44)* to fit the sub-band radio photometry with a synchrotron spectral break model [*(60)*, their equation 1]. We adopt top-hat priors over wide ranges in all parameters and find that each radio epoch is well fitted by this model (fig. 1). We are primarily interested in the observed peak frequency $\nu_{peak}$ and flux $F_{peak}$ rather than the extrapolated values given by the formula, because they allow us to measure the radius *R* and magnetic field *B* of the shock *(15)*. To estimate the distribution of these quantities, we use the spectral break equation to generate a finely resolved model spectrum for each sample of model parameters in the posterior distribution and read off the peak values from each model spectrum *(38)*. The peak frequencies and fluxes are listed along with their uncertainties in table S3, and have statistical uncertainties of less than 1% in both epochs. The dominant sources of uncertainty in the derived shock properties are systematic, due to the unmeasured parameters $\epsilon_e$, $\epsilon_B$, and $f$ described below.

Shock radius and magnetic field

The frequency and luminosity of a shock's synchrotron self-absorption peak can be used to estimate the shock's radius *R* and magnetic field *B.* Assuming an optically thin spectral index of $\alpha_{thin} = 1$ (appropriate for VT 1210+4956; see table S3), we calculate $R$ and $B$ using equations *(15)* in units scaled roughly to VT 1210+4956.

$$R = 7.5 \times 10^{16} \left(\frac{\epsilon_e}{\epsilon_B}\right)^{-\frac{1}{19}} \left(\frac{f}{0.2}\right)^{-\frac{1}{19}} \left(\frac{L_p}{10^{29} \text{ erg/s/Hz}}\right)^{\frac{9}{19}} \left(\frac{\nu_p}{5 \text{ GHz}}\right)^{-1} \text{ cm} \qquad \text{(S1)}$$

$$B = 0.21 \left(\frac{\epsilon_e}{\epsilon_B}\right)^{-\frac{4}{19}} \left(\frac{f}{0.2}\right)^{-\frac{4}{19}} \left(\frac{L_p}{10^{29} \text{ erg/s/Hz}}\right)^{-\frac{2}{19}} \left(\frac{\nu_p}{5 \text{ GHz}}\right) \text{ G} \qquad \text{(S2)}$$

where $\epsilon_B$ and $\epsilon_e$ are the fraction of shock energy in the magnetic field and in relativistic (energy > 511 keV) electrons respectively, $f$ is the fraction of the spherical volume of radius $R$ that is emitting the synchrotron emission, and $L_p$ and $\nu_p$ are the luminosity and frequency of the radio spectrum's spectral peak. The strongest dependencies in these equations are on $L_p$ and $\nu_p$, which are directly measured from the radio spectrum (table S3). Both $R$ and $B$ are weakly dependent on the unmeasured quantities $\frac{\epsilon_e}{\epsilon_B}$, and $f$. We adopt a similar convention to studies of other non-relativistic radio supernovae in assuming $\epsilon_e = \epsilon_B = 0.1$ [see *(17)* for a discussion]. Because of the likely aspherical CSM geometry, we marginalize over $f$ by assuming it is drawn from a uniform distribution between 0.1 and 0.5 *(38)*. The resulting values are listed in table S4.

Energy in the shock

The energy $U$ dissipated in the shock is *(17)*:

$$U = \left(\frac{1}{\epsilon_B}\right) \frac{4}{3} \pi f R^3 (B^2/8\pi) \qquad \text{(S3)}$$

$$= 2.7 \times 10^{49} \left(\frac{\epsilon_B}{0.1}\right)^{-1} \left(\frac{\epsilon_e}{\epsilon_B}\right)^{-\frac{11}{19}} \left(\frac{f}{0.2}\right)^{\frac{8}{19}} \left(\frac{L_p}{10^{29} \text{ erg/s/Hz}}\right)^{\frac{23}{19}} \left(\frac{\nu_p}{5 \text{GHz}}\right)^{-1} \text{ erg} \qquad \text{(S4)}$$

We make the same assumptions for $\epsilon_e, \epsilon_B$, and $f$ as before; the dependence on these parameters is much stronger for the energy. The largest systematic uncertainty arises from the factor ($1/\epsilon_B$), with all other uncertainties likely amounting to less than a factor of 2 in the energy. In gamma ray bursts, $\epsilon_B$ has been estimated to be as small as ~$10^{-6}$, however, for non-relativistic radio supernovae, it is often assumed to be between 0.1 - 0.5 *(14)*. For our comparison with other radio supernovae in fig. 4, we normalize all values to $\epsilon_B = 0.1$. Under this assumption, we find an energy of $(5.4^{+0.9}_{-1.2}) \times 10^{49}$ erg in follow up epoch 1 (table S4). If instead we assume that $\epsilon_B$ is drawn from a log-uniform distribution between $10^{-3}$ and 0.1, we find a ~3x larger value of $(14^{+13}_{-7}) \times 10^{49}$ erg *(38)*.

Radio shock velocity and CSM density

By measuring the shock radius $R$ at two epochs, using identical values for $\epsilon_e, \epsilon_B$, and $f$ for both epochs in each instance of the Monte Carlo simulation, we can determine the average shock velocity between the two: $v_{radio} = \Delta R_{Epoch1-2}/\Delta t_{Epoch1-2}$. We find a shock velocity for VT 1210+4956 of 1780 ± 290 km s$^{-1}$. If we adopt the wide range of $\epsilon_B$ described above, we find a value of $1570^{+290}_{-270}$ km s$^{-1}$. Both values are similar to the FWHM velocity of the H$\alpha$ line, $v_{H\alpha} = 1345^{+64}_{-56}$ km s$^{-1}$, consistent with our interpretation of the H$\alpha$ emission coming from a dense shell behind the forward shock *(61)*. Because the H$\alpha$ velocity is more tightly constrained than the radio velocity, we use its value as a proxy for the shock velocity in estimating the CSM density: $v_{shock} \equiv v_{H\alpha} \approx v_{radio}$. Using measurements of the magnetic field and the shock velocity, we can estimate the density of the gas swept up between the follow-up epochs. Applying the shock jump conditions in the limit of a strong shock, we derive relations $P_1 + \rho_1 v_1^2 = P_2 + \rho_2 v_2^2$ and $\frac{\rho_2}{\rho_1} = \frac{v_1}{v_2} \approx 4$, where $P$ is the pressure, $v$ is the forward shock velocity, $\rho$ is the mass density, and the subscripts 1 and 2 refer to pre- and post-shock quantities respectively. The post-shock (thermal) pressure is given by $P_2 = \frac{1}{\epsilon_B}\frac{B^2}{8\pi}$. Material is swept up by the shock at a speed of $v_1 \approx v_{shock}$ and the mass density is related to the number density by $\rho = \mu m_p n$, where $\mu$ is the mean atomic mass. Combining these relations, the pre-shock CSM density is:

$$n_1 = \frac{4}{3\mu m_p} \frac{(1/\epsilon_B)(B^2/8\pi)}{v_{shock}^2} \tag{S5}$$

Scaled to typical values for the VLASS transients and assuming a pure hydrogen gas ($\mu = 1$), the pre-shock density is

$$n_1 = 3.9 \times 10^6 \left(\frac{\epsilon_B}{0.1}\right)^{-1} \left(\frac{\epsilon_e}{\epsilon_B}\right)^{\frac{-8}{19}} \left(\frac{f}{0.2}\right)^{\frac{-8}{19}} \left(\frac{L_p}{10^{29}\text{erg/s/Hz}}\right)^{\frac{-4}{19}} \left(\frac{v_p}{3\text{GHz}}\right)^2 \left(\frac{v_{shock}}{1000\text{km/s}}\right)^{-2} \text{cm}^{-3} \tag{S6}$$

Assuming that $\epsilon_B = \epsilon_e = 0.1$, $f$ is uniformly distributed between 0.1 and 0.5, and adopting the H$\alpha$ velocity as discussed above, we find a density of $n_1 = 1.1^{+0.3}_{-0.2} \times 10^6 \text{cm}^{-3}$. Allowing $\epsilon_B$ to vary in a wide range as with the energy changes this estimate to $n_1 = 4^{+7}_{-2} \times 10^6 \text{cm}^{-3}$ *(38)*. The resulting density of ~$10^6$ cm$^{-3}$ is higher than the densities observed in non-nuclear regions of galaxies, with the exception of giant molecular cloud (GMC) cores and extreme stellar outflows *(25)*. This high gas density is located within a radius of 0.05 pc of a highly energetic explosion associated with an evolved star forming region. Within this radius, GMC cores would be disrupted through feedback within the lifetime of any massive star *(62)*. We conclude that this mass was deposited through a pre-explosion outflow from the progenitor.

Total shocked mass in the circumstellar shell

From the inferred densities, velocities and shock radii, we set a lower limit on the mass swept up within the shock. The total mass swept up in the $\Delta t$ ~1 year between our two radio follow-up epochs is $M_{shocked} \approx 4\pi R^2 f_{shell}(v_{shock}\Delta t) m_p n_1 = 0.8 M_\odot (f_{shell})$, where $f_{shell}$ is the fraction of the shocked shell that is emitting radio emission, and $m_p$ is the mass of an ionized hydrogen atom. The radio component of the mass estimate scales with the density and can thus be scaled downwards by ~3x or upwards by 10x for the range of $\epsilon_B$ values considered above. The gas producing the radio emission is shock heated to a temperature of $10^7$ K, a temperature at which hydrogen recombination is too inefficient to produce the H$\alpha$ line. Thus, there is an additional cooler component of gas at a temperature where recombination is efficient, providing the H$\alpha$ emission associated with the shock. We estimate the mass in this phase using the luminosity of the H$\alpha$ line, which scales inversely with the density. The total mass of this cool ionized gas is given by $M_{ionized} = \frac{m_p L_{H\alpha}}{(\alpha_{B,H\alpha}) n_{1}(hc/6563 \text{Å})}$, where $\alpha_{B,H\alpha}$ = 1.17×10$^{-13}$ cm$^3$ s$^{-1}$ is the effective Case B recombination coefficient for H$\alpha$ at a temperature of $10^4$ K *(26)*, and $L_{H\alpha}$ is the H$\alpha$ luminosity, and $hc$ is the product of Planck's constant and the speed of light. Assuming that the H$\alpha$ producing gas has a similar density to the pre-shock gas, we find an ionized mass of ~$0.7 M_\odot$ for VT 1210+4956. Combining the two phases, we find that the minimum shell mass (responsible for radio and optical emission at the time of our observations) is ~$0.8 M_\odot$.

Mass loss rates and CSM velocities

The mass density $\rho$ at a given radius $R$ is related to the pre-explosion mass loss rate $\dot{M}$ and CSM velocity $v$ by $\dot{M}/v = 4\pi R^2 \rho$, assuming constant $\dot{M}$ and $v$ over the scale probed by a given observation. For a pure hydrogen CSM, we have:

$$\dot{M} = 3.3 \times 10^{-2} \left(\frac{n}{10^6 \text{cm}^{-3}}\right) \left(\frac{R}{10^{17} \text{cm}}\right)^2 \left(\frac{v}{100 \text{ km/s}}\right) M_\odot \text{ yr}^{-1} \qquad (S7)$$

For VT 1210+4956, we find a mass loss rate of $\frac{\dot{M}}{(v/100 \text{ km s}^{-1})} = 4 \times 10^{-2} M_\odot$ yr.

Mass lost from stars has a terminal velocity similar to the escape velocity of the star. These velocities range from 10s of km/s for systems with loosely bound atmospheres such as giant stars and common envelopes, to ~1000 km/s for Wolf Rayet stars *(27)*. The densest observed winds tend to move at velocities closer to ~50 km/s, given the additional luminosity needed to drive the extreme mass loss *(27)*. Given that our observed mass loss rates are higher, we take this as a rough lower bound on the mass loss velocity in wind scenarios. For binary mass loss, the same escape velocity principle applies. Given that the majority of interactions occur on the main sequence or during first expansion to a red giant, we adopt the same fiducial lower bound on the mass loss velocity. In type IIn supernovae, the narrow H$\alpha$ line indicates the velocity of the pre-explosion mass loss. The LRIS velocity resolution sets an upper limit of ~270 km/s. With the velocity range of 50-270 km/s, the gas accounting for the massive shell we infer must have been lost ~100-1000 years prior to explosion.

Constraints on the source of MAXI 140814A

MAXI 140814A has a spectrum that is much softer than that of gamma-ray bursts. It is more reminiscent of SN 2008D, which has been interpreted as a sub-relativistic shock breakout from an optically thick stellar wind *(23)*. However, MAXI 140814A's luminosity is three orders of magnitude brighter than the SN 2008D X-ray flare *(23)*.

We first assume that the source is a spherical sub-relativistic stellar explosion, with velocity $\beta c$, where $c$ is the speed of light, and scale to a total burst energy $E \approx 3 \times 10^{47}$ erg, a burst duration $t \approx 15$ s, and a typical photon energy $\epsilon_\gamma \approx 5$ keV. The radiated energy in such explosion from any fluid element is at most equal to its kinetic energy, so the mass of the source must satisfy:

$$m \gtrsim \frac{2E}{\beta^2 c^2} = 3 \times 10^{-7} M_\odot \frac{E}{3 \times 10^{47} \text{erg}} \beta^{-2} \tag{S8}$$

The burst duration sets an upper limit on the source radius, $R < 6.5 R_\odot (t/15s)$. Thus the optical depth of the source must satisfy

$$\tau \approx \kappa \frac{m}{4\pi R^2} \gtrsim 250 \frac{E}{3 \times 10^{47} \text{ erg}} \frac{\kappa}{0.2 \text{ cm}^2 \text{ g}^{-1}} \left(\frac{t}{15s}\right)^{-2} \beta^{-2} \tag{S9}$$

where $\kappa$ is the opacity of the source material per unit of mass. For MAXI 140814A, $\tau \gg \beta^{-1}$, which implies that the radiation is trapped in the source and cannot diffuse to the observer over the observed duration. Thus, the source cannot be non-relativistic and spherical. For an aspherical non-relativistic source, the optical depth is higher for the same amount of mass, and so is the diffusion time unless the outflow geometry is finely tuned. Therefore, we conclude that the outflow is relativistic.

A relativistic explosion solves the high diffusion time problem. A relativistic spherical source that is shocked to a Lorentz factor $\gamma$, has an internal energy $E \approx mc^2\gamma(\gamma - 1)$ and therefore,

$$m \gtrsim \frac{E}{\gamma(\gamma-1)c^2} = 1.5 \times 10^{-7} M_\odot \frac{E}{3 \times 10^{47} \text{erg}} \left(\gamma(\gamma-1)\right)^{-2} \tag{S10}$$

at the same time, relativistic effects dictate $R < 2ct\gamma^2 = 13 R_\odot (t/15s)\gamma^2$, so

$$\tau \approx \kappa \frac{m}{4\pi R^2} \gtrsim 125 \frac{E}{3 \times 10^{47} \text{erg}} \left(\frac{t}{15s}\right)^{-2} \left(\frac{\kappa}{0.2 \text{ cm}^2 \text{ g}^{-1}}\right)^{-2} \gamma^{-4}(\gamma-1)^{-2} \tag{S11}$$

The requirement $\tau < \beta^{-1}$ implies $\gamma > 2.5$.

However, a relativistic spherical explosion is also inconsistent with the data. First, accelerating enough mass to relativistic velocities in a spherical explosion of a star requires an extreme energy of ~$10^{53}$ erg *(63)*. Second, if the stellar envelope is at rest at the time that the explosion energy is released (i.e., there is no double explosion) the photons that are generated behind the shock are radiated upon the breakout of the shock and their typical observed energy is 100 keV *(63)*, higher than observed.

We conclude that the most likely explanation for MAXI 140814A is a non-spherical relativistic expansion, such as a jet. A collimated outflow does not change the lower limit on the source Lorentz factor derived above. If the radiation of the source is beamed towards the observer (i.e., the angle between the source velocity and the line of sight is smaller than $1/\gamma$) the limits are similar to the spherical case, and if the radiation is beamed away from the observer the limits are more severe *(64)*.

**Supplementary Text**

Consistency checks

Other than requiring that high-energy transients occur before the first radio observation of VT 1210+4956, we did not explicitly consider the inferred explosion date in the FAP calculation. However, the timing of GRB 140814A is consistent with the evolution of the radio spectrum. When supernova ejecta begin to interact with a dense CSM, the forward shock velocity decreases suddenly, due to the large density jump *(65)*. The shock velocity is then reaccelerated by the pressure produced by the shocked supernova ejecta. The properties of the ejecta and CSM can thus be used to estimate the forward shock velocity as a function of time. Assuming an ejecta mass of ~$2 M_\odot$ and kinetic energy of ~$10^{51}$ erg, similar to SN 2014C *(28)* and SN 2001em *(31)*, we find that the forward shock velocity reaches ~2000 km/s approximately 3 years after the explosion *(65)*. We assume a CSM shell of uniform density ~2 x $10^6$ cm$^{-3}$ with total mass ~1$M_\odot$ at radius ~9 x $10^{16}$ cm, as inferred from the radio analysis (table S4, see discussion above). This is consistent with the ~3.5 year delay between GRB 140814A and the epoch at which we infer a radio-based shock velocity of ~1800 km/s for VT 1210+4956.

The column density of hydrogen, $N_H$, in a spherical CSM shell with the assumed values is $N_H$ ~ 2 x $10^{22}$ cm$^2$, at which the optical depth at 2 keV is ~1, rapidly decreasing at higher energies. The column density would be lower for an off-axis line of sight in a toroidal CSM geometry. This is consistent with GRB 140814A's observed soft X-ray spectrum being intrinsic rather than a product of obscuration.

We considered other potential sources of short, soft X-ray transients. One such class is a shock breakout from a supernova similar to the one observed in SN 2008D, which was 3 orders of magnitude less luminous than MAXI 140814A *(23)*. For MAXI to detect such a breakout, it would have to be at a distance within ~5 Mpc. At that distance, the contemporaneous All Sky Automated Survey for Supernovae (ASASSN) *(66)* was complete to supernovae with a V or g band absolute magnitude brighter than -12, easily sufficient to detect core collapse and stripped envelope supernovae months after explosion *(67–69)*. There are no cataloged supernovae that are located within the variable source region and exploded within ~1 month of the X-ray transient *(49)*. This rules out shock breakout as an explanation for the MAXI transient. At the ~149 Mpc distance of VT 1210+4956, the ASASSN limiting magnitude of > –20.5 is not constraining for a wide range of supernova types *(67–69)*.

Another potential class is a flare from a Galactic source, such as a star or X-ray binary. MAXI 140814A is located at a Galactic latitude of +66 deg, where the column density of Galactic sources is low. In a search of the SIMBAD database *(70)*, we did not find any known flare stars or X-ray binaries within the MAXI localization region.

X-ray flares from stars generally last from minutes to hours, as opposed to ~15s for MAXI 140814A *(71)*. The peak luminosity of these flares is generally 1-2 orders of magnitude brighter than the quiescent X-ray luminosity from the star. To search for a potential quiescent counterpart, we checked the Röntgensatellit (ROSAT) all sky catalog *(72)* and found that the brightest X-ray source within a 3-degree radius of the MAXI 140814A constant flux localization has a ROSAT count rate of 0.2 counts/s in the 0.1-2.5 keV band. This corresponds to a flux < 1000x the 2-4 keV flux of MAXI 140814A for a range of power law and blackbody models, further disfavoring a flare star origin.

X-ray binaries are known to brighten by several orders of magnitude when transitioning between the low-hard state and high-soft state. These transitions typically last for days to months *(73)*, whereas GRB 140814A was detected for ~15s in only one 92-minute orbit of the ISS. Luminous shorter timescale flares have been observed from accreting compact object systems, with some burst-only systems displaying quiescent fluxes > 3 orders of magnitude lower than the peak of the bursts *(74)*. The vast majority of known bursters have a Galactic latitude between -15 and +15 deg *(75)* in contrast to the +66.4 deg Galactic latitude of GRB 140814A. Additionally, these bursts are observed to, and theoretically expected to repeat *(76, 77)*. In particular, bursts with durations of 10-100s have typical recurrence timescales of hours to days *(77)*. No other bursts from the localization region of GRB 140814A have been reported from X-ray follow-up observations of the field, or the >11 years of observations from the MAXI GSC.

AGN are known to display variability at all wavelengths, and have been observed to launch new jets *(78)*. However, it is difficult to explain a 15s X-ray flare from an AGN. The characteristic X-ray variability timescale for an AGN scales with its supermassive black hole mass, $M_{BH}$, at 5 days × $(M_{BH} / 10^8 M_\odot)^2$ *(79)*. Thus, the characteristic black hole mass for a variability timescale of ~1 minute is ~$10^4 M_\odot$. The peak X-ray luminosity of MAXI 140814A is ~4 x $10^{46}$ erg/s × $(d/150 Mpc)^2$, which corresponds to ~$10^4$ times the Eddington luminosity for a $10^4 M_\odot$ black hole at distance $d$ = 150 Mpc. A luminosity at the Eddington limit would require an intermediate-mass black hole at ~1.5 Mpc; no such sources are known. There are also no catalogued galaxies within the MAXI variable source localization region closer than 10 Mpc *(80)*.

Constraints on an optical supernova

Due to its proximity to the Sun, the optical sky survey coverage of the location of VT 1210+4956 at the time of MAXI 140814A was sparse. Nevertheless, available observations can constrain the possibility of a supernova at the position of VT 1210+4956 in the months following MAXI 140814A.

The only optical constraint reported in the literature comes from the Mobile Astronomical System of Telescope Robots (MASTER) II telescope in Kislovodsk, which followed up a region that likely contains VT 1210+4956 *(52)* starting 10.3 hours after the MAXI transient time. The upper limit from this observation is an apparent magnitude of >18.5 in single 180s frames and >20.0 in 10 images co-added. The latter limit corresponds to an absolute magnitude of -15.9 at the distance of VT 1210+4956. The Balmer decrement of 4.9 from the ratio of the narrow Hα and Hβ lines in the 2018 optical spectrum of VT 1210+4956 implies a reddening E(B-V) = 0.46. Assuming a standard reddening curve *(81)*, this corresponds to an extinction of 0.8 – 2.8 magnitudes in the MASTER II bands (B, V, R, and I), making the effective limit somewhat shallower ($\gtrsim$ -16.7 to -18.7). Due to the rise time of days to weeks for all but the fastest supernovae *(18)* this limit does not rule out most types of supernova.

We measured two additional constraints using data from the Palomar Transient Factory (PTF) *(82)* and the Panoramic Survey Telescope and Rapid Response System (Pan-STARRS) *(83)*. The PTF observations were taken in g band ~2.2 months after GRB 140814A in October 2014. There are no g band PTF observations of the field prior to this date, so we convolved the PTF image to the resolution of a pre-explosion Pan-STARRS image and subtracted the two using a difference imaging pipeline *(84)*. We do not detect a source in the difference image to limits of g > 19.6, g > 20.0, and g > 20.4 on 2014 October 19, 20, and 23 respectively. The deepest of these limits corresponds to an absolute magnitude G > -15.5, or an extinction corrected limit of G > -18.1. The Pan-STARRs observations were taken in I band on 2015 January 12. We stacked all available images from this epoch and used the pipeline *(84)* to subtract them from a stack of Pan-STARRS I band images from 2014 May 9. We do not detect any transient or variable emission to a limiting apparent magnitude of I > 21.1, corresponding to an absolute magnitude of -14.8 and an extinction corrected value of I > -15.6. The PTF and Pan-STARRS limits rule out superluminous supernovae (e.g. *85*), and luminous type IIn supernovae with large amounts of mass lost close to explosion (*69*,*86*). However, they are not constraining for other types of stripped envelope supernovae (*68*), including analogs of SN 2014C (*28*) and PTF 11qcj (*19*). This is consistent with the early time counterpart of VT 1210+4956 having been a stripped envelope supernova, though does not require this.

Free-free absorption and asymmetric CSM

In our radio follow-up observations of VT 1210+4956, we measured peaks in the broadband radio spectra at $5.06 \pm 0.02$ GHz in the first epoch and $4.49 \pm 0.03$ GHz in the second epoch (fig. 1). In the previous sections, we assumed that this peak was due to synchrotron self-absorption, leading to results that are consistent with the optical spectrum of VT 1210+4956. Here we test whether the spectra could instead be explained with free-free absorption.

The free-free optical depth $\tau_\nu$ of radio emission passing through a plasma is given by:

$$\tau_\nu = 3.28 \times 10^{-7} \left(\frac{\nu}{\text{GHz}}\right)^{-2.1} \int_{LOS} \left(\frac{T}{10^4\,\text{K}}\right)^{-1.35} \left(\frac{n_e}{\text{cm}^{-3}}\right)^2 \frac{ds}{\text{pc}} \qquad (S12)$$

where $\nu$ is the frequency of light being absorbed/transmitted, and $T$ and $n_e$ are the electron temperature and density in a region of length $ds$ along the line of sight (LOS) (*87*). The optical depth scales as $\tau_\nu \propto \nu^{-2.1}$, which implies that on the optically thick side, the flux density scales super-exponentially with frequency as $S_\nu \propto \nu^{-1} \exp(-\tau_\nu) \propto \nu^{-1} \exp(-\nu^{-2.1})$, where the additional factor of $\nu^{-1}$ arises from the intrinsic spectral index of the emission, which is ~1 (Table S3).

Combining this relation with our lowest frequency detection of $0.56 \pm 0.07$ mJy at 1.14 GHz, we set an upper limit on the optical depth to free-free absorption at the $5.87 \pm 0.02$ mJy peak at 5.06 GHz in the first epoch. If there is no synchrotron self-absorption, all of the drop in flux from the peak at $S_5 = 5.87$ mJy to the low frequency measurement of $S_1 = 0.56$ mJy is due to free-free absorption. In that case, we have that $S_5/S_1 = (5.06/1.14)^{-1} \exp\left(-(\tau_{5,ff} - \tau_{1,ff})\right)$, where $\tau_{5,ff}$ and $\tau_{1,ff}$ are the free-free optical depth at 5.06 and 1.14 GHz respectively. This implies that $\tau_{1,ff} - \tau_{5,ff} = 3.9$. Simultaneously, we find that $\tau_{1,ff}/\tau_{5,ff} = \left(\frac{1.14}{5.06}\right)^{-2.1} \sim 22.9$. This then implies that $\tau_{5,ff} = 0.17$. Thus, the maximum possible attenuation from free-free absorption at the 5 GHz peak is $1 - e^{-0.17} = 16\%$, with any contribution from synchrotron self-absorption reducing this even closer to 0.

We improve this limit by attempting to reproduce the detailed shape of the observed peak with free-free absorption. To do so, we assume that a synchrotron source with unabsorbed spectrum $S_\nu = A\nu^\alpha$ (where $\alpha$ is the spectral index and $A$ is a scale factor) is being attenuated by a slab of material with constant temperature $T$ and emission measure $EM = \int_{LOS} \left(\frac{n_e}{cm^{-3}}\right)^2 \frac{ds}{pc}$. We used EMCEE to explore the parameter space of *EM*, *T*, *A*, and *α*, assuming uninformative priors for all parameters *(39)*. This model can approximate the radio spectrum at frequencies > 2 GHz. However, at lower frequencies, the super-exponential decay in flux becomes dominant. At 1.5 GHz, the predicted flux drops to 0.359 ± 0.006 mJy, nearly 4x lower than our measured value of 1.22 ± 0.04 mJy. At 1.14 GHz, the model predicts 0.0169 ± 0.0007 mJy, over 30x lower than our measured value of 0.56 ± 0.07 mJy. This confirms that the peak cannot be explained by free-free absorption.

The lack of free-free absorption requires an explanation, given the high CSM density. Above we estimated the density of the CSM to be ~2.2 x $10^6$ cm$^{-3}$. This density depends primarily on the shock velocity (which is independently and consistently measured using both the Hα line width and the change in shock radius), the peak frequency (which has uncertainty less than 1%), and the assumed value of 0.1 for the magnetic field energy fraction $\epsilon_B$ (which can vary, but cannot decrease the density by more than a factor of ~3). If we approximate the CSM between the first and second epochs as a spherical shell with uniform density and temperature, the free-free optical depth at 5 GHz is $\tau_\nu \sim 88(T/10^4 K)^{-1.35} (n_e/2.2 \times 10^6 cm^{-3})^2 \gg 1$ for the estimated density at typical H II region temperatures. Any dense ionized gas that has yet to be shocked along the line of sight would only increase the free-free absorption.

To reconcile the inferred lack of free-free absorption at 5 GHz with the implied density, we require either a high temperature in the pre-shock CSM, a lower density along the line of sight, or both. A high temperature, driven by Compton heating from the central X-ray source, was invoked to explain a similar phenomenon in the fast blue optical transient AT 2018cow *(17)*. However, the X-ray flux of this source faded by a factor of ~$10^3$ within the first 100 days of discovery *(88)*. An analogous source would have to be substantially longer-lived to maintain a temperature necessary to suppress the free-free absorption, or would have to be powerful enough at early times to heat the dense CSM to $\gg 10^7$ K where it would not have enough time to cool to the typical H II region equilibrium temperature of $10^4$ K in the years before encountering the shock. A simpler solution is if the CSM is asymmetric, with the line of sight being along a lower density direction. Asymmetric mass loss has been predicted for binary interactions. Hydrodynamic simulations have found that the CSM from binary interaction generally forms a toroidal geometry that can be orders of magnitude denser in the midplane than along the poles *(4, 7)*. If our line of sight is not oriented directly along the midplane, it would be through a substantially smaller column density of gas. Much of that gas may already be shocked, because shocks propagate faster through lower density regions. For a shock velocity $v_{shock}$, the pre-shocked gas would be heated to a temperature of $T = \frac{3}{16} \frac{m_p}{k_B} (v_{shock})^2 \sim 5.7 \times 10^8 K \left(\frac{v_{shock}}{5000 km/s}\right)^2$, far above the ~$10^5$ K temperature required to suppress free-free absorption.

Rate of similar radio transients

VT 1210+4956 was identified in a search for distance $d < 200$ Mpc off-nuclear extragalactic transients with a flux-limited detection limit of ~0.7 mJy. With its 3 GHz luminosity at the detection epoch of $7.1 \times 10^{28}$ erg/s, it would have been detectable out to 291 Mpc. Its host galaxy has an r-band apparent magnitude of 17.7 at a distance of 150 Mpc, and would be easily detectable at 200 Mpc. Likewise, the broad emission feature would be detectable by the same follow-up Keck observation at 200 Mpc. This leaves only one major source of incompleteness: the 6195 deg$^2$ overlap between VLASS Epoch 1.1 and FIRST. The expected all-sky number of 3 GHz transients at this radio luminosity is thus ~6.7 within 200 Mpc, corresponding to ~2 x 10$^{-7}$ transients Mpc$^{-3}$. To convert this density to a volumetric rate, we require the duration for which each event would be detectable in our search. The lower limit is set by the time between the first detection in VLASS Epoch 1.1 and the latest detection in VLASS Epoch 2.1, spanning ~2.5 years (table S2). Given the unknown CSM profile outside the current radius reached, it is unclear how long it will take before the source fades below our detection limit. However, there is a maximum effective timescale probed by our search: the time between non-detection in FIRST and detection in VLASS, which is ~7500 days for VT 1210+4956. Sources that are older than this will appear as variable sources rather than transients, and thus are not included in our sample. Using this range of detectable ages, we estimate a range of volumetric rates for similarly luminous transients of ~(1–8) x 10$^{-8}$ transients Mpc$^{-3}$ yr$^{-1}$.

**Table S1. Best fitting emission line model components for VT 1210+4956.**
Based on the Keck/LRIS spectrum on 2018 April 13 (0.37 years after detection in VLASS). Columns are the name of the emission line component, the rest wavelength of the line $\lambda_{rest}$, the best-fitting central wavelength of the line component $\lambda_{fit}$, the velocity full-width-half-maximum (FWHM) of the line component, the total flux of the component, and its luminosity. Upper and lower uncertainties are respectively the 86th and 14th percentiles of the posterior probability distribution.

| Line | $\lambda_{rest}$ (Å) | $\lambda_{fit}$ (Å) | FWHM (km/s) | Integrated flux ($10^{-17}$ erg s$^{-1}$ cm$^{-2}$) | Luminosity ($10^{36}$ erg s$^{-1}$) |
|---|---|---|---|---|---|
| Hα broad | 6562.8 | $6559.6^{+0.49}_{-0.45}$ | $1345^{+64}_{-56}$ | $27.7^{+1.1}_{-0.9}$ | $727^{+29}_{-25}$ |
| Hα narrow | 6562.8 | | < 270 | $10.7^{+0.5}_{-0.4}$ | $282^{+13}_{-11}$ |
| NII 6583 Å | 6583.45 | | < 270 | $2.3^{+0.3}_{-0.3}$ | $61^{+7}_{-7}$ |
| NII 6548 Å | 6548.05 | | < 270 | $0.9^{+0.3}_{-0.3}$ | $23^{+8}_{-8}$ |

**Table S2. Radio fluxes from single-band images of VT 1210+4956.**
Columns are the observation date, the fluxes $S_\nu$ in VLA bands centered at frequency $\nu$, and the name of the observation epoch. Observation dates are given relative to detection in VLASS of VT 1210+4956 on 2017 November 20. Uncertainties are the RMS noise in a nearby patch with no emission. The upper limit from FIRST is 3x the RMS noise. There is an additional ~5% systematic uncertainty in the overall flux density scale, which is correlated between bands and constant between epochs, as we used the same calibrators. There is an additional uncertainty of ~20% in fluxes from VLASS quick-look imaging *(9)*.

| Time relative to VLASS detection (years) | $S_{1.5GHz}$ (mJy) | $S_{3GHz}$ (mJy) | $S_{6GHz}$ (mJy) | $S_{10GHz}$ (mJy) | $S_{15GHz}$ (mJy) | Epoch |
|---|---|---|---|---|---|---|
| -20.59 | < 0.41 | | | | | FIRST |
| 0 | | 2.8 ± 0.1 | | | | VLASS 1.1 |
| 0.5233 | 1.22 ± 0.04 | 4.54 ± 0.02 | 6.11 ± 0.01 | 5.22 ± 0.01 | 3.85 ± 0.01 | Follow up 1 |
| 1.441 | 1.3 ± 0.3 | 4.13 ± 0.05 | | | | Follow up 2 |
| 1.463 | | | 4.62 ± 0.01 | 3.04 ± 0.02 | | Follow up 2 |
| 2.454 | | 4.0 ± 0.2 | | | | VLASS 2.1 |

**Table S3. Fitted properties of VT 1210+4956's radio spectrum from follow-up observations with the VLA.**

The columns are the epoch (same definition as table S2), the spectral peak frequency $\nu_{peak}$, the flux at the peak frequency $S_{\nu_{peak}}$, and the asymptotic optically thin spectral index $\alpha$. Upper and lower uncertainties are the 86th and 14th percentiles of the posterior probability distribution. We jointly fit the two observations in follow up epoch 2.

| Time relative to VLASS detection (years) | $\nu_{peak}$ (GHz) | $S_{\nu_{peak}}$ (mJy) | $\alpha$ |
|---|---|---|---|
| 0.5233 | $5.06^{+0.03}_{-0.02}$ | $5.87^{+0.02}_{-0.02}$ | $-1.04^{+0.02}_{-0.02}$ |
| 1.452 | $4.48^{+0.03}_{-0.02}$ | $5.10^{+0.02}_{-0.02}$ | $-1.05^{+0.11}_{-0.17}$ |

**Table S4. Derived properties of the VT 1210+4956 shock and the CSM in which it is propagating.**

The columns are the epoch of the observation (same definition as table S2), the shock radius $R$, the shock magnetic field $B$, the energy dissipated within the shock $U$, the pre-shock CSM density $n$, and the assumed distributions of microparameters $\epsilon_e$, $\epsilon_B$, and $f$. For the microparameters assumed, A denotes $\epsilon_e = \epsilon_B = 0.1$, with $f$ drawn from a uniform distribution between 0.1 and 0.5, while B denotes the same distributions for $\epsilon_e$ and $f$ with $\epsilon_B$ drawn from a log-uniform distribution between $10^{-3}$ and 0.1, (see text). The uncertainties are the 86th and 14th percentiles of the posterior probability distribution. Values between epochs are positively correlated assuming no evolution in $\epsilon_e$, $\epsilon_B$, and $f$, with an implied velocity from the change in $R$ between the two epochs of $1780^{+290}_{-290}$ km s$^{-1}$ for microparameter set A, and $1570^{+290}_{-270}$ km s$^{-1}$ for microparameter set B (see text).

| Time relative to VLASS detection (years) | $R$ ($10^{16}$ cm) | $B$ (mG) | $U$ ($10^{49}$ erg) | $n$ ($10^6$ cm$^{-3}$) | Set of assumed microparameter distributions |
|---|---|---|---|---|---|
| 0.5233 | $8.9^{+0.3}_{-0.2}$ | $392^{+5}_{-2}$ | $5.4^{+0.9}_{-1.2}$ | $1.1^{+0.3}_{-0.2}$ | A |
| 0.5233 | $7.9^{+0.6}_{-0.6}$ | $246^{+10}_{-7}$ | $14^{+13}_{-7}$ | $4^{+7}_{-2}$ | B |
| 1.452 | $9.4^{+0.3}_{-0.2}$ | $353^{+5}_{-3}$ | $5.1^{+0.9}_{-1.1}$ | | A |
| 1.452 | $8.4^{+0.7}_{-0.7}$ | $221^{+8}_{-6}$ | $13^{+12}_{-6}$ | | B |

**Table S5. Energies, velocities, radii, densities, and references for sources plotted in Fig. 4.**
Columns *U*, *R*, and *n* are as defined in table S4 and are either calculated from radio data presented in references assuming $\epsilon_e = \epsilon_B = 0.1$ (see text) or taken directly from the reference(s). Dimensionless shock velocities $\Gamma\beta$, where $\Gamma$ is the bulk Lorentz factor and $\beta$ is the velocity divided by the speed of light, are either taken from the reference(s) or estimated from *R* assuming a constant velocity since explosion.

| Object(s) | $U$ ($10^{49}$ erg) | $\Gamma\beta$ ($10^{-2}$) | $R$ ($10^{16}$ cm) | $n$ ($10^6$ cm$^{-3}$) | Reference(s) |
|---|---|---|---|---|---|
| SN 1978 K | 0.36 | 2.0 | 12 | $9.8 \times 10^{-3}$ | (15) |
| SN 1979C | $1.4 \times 10^{-2}$ | 1.2 | 1.4 | 1.1 | (15) |
| SN 1980 K | $2.4 \times 10^{-4}$ | 0.63 | 0.27 | 7.4 | (15) |
| SN 1986 E | $2.7 \times 10^{-4}$ | 0.47 | 0.31 | 13 | (15) |
| SN 1986J | 0.10 | 0.97 | 0.32 | 1.1 | (15) |
| SN 1988Z | 0.20 | 1 | 3.9 | 0.94 | (15) |
| SN 1990B | $3.2 \times 10^{-2}$ | 13 | 3.0 | $1.1 \times 10^{-3}$ | (15) |
| SN 1993J | $1.9 \times 10^{-2}$ | 3.3 | 1.3 | 0.15 | (15) |
| SN 1994I | $1.6 \times 10^{-2}$ | 13 | 1.2 | $1 \times 10^{-2}$ | (15) |
| AT 2018cow | 0.74 | 13 | 0.73 | 1.3 | (17) |
| SN 2009bb | 0.27 | 70 | 5 | $2.4 \times 10^{-5}$ | (22) |
| SN 2014C (shell) | $6.5 \times 10^{-2}$ | 0.5 | 5.5 | 0.55 | (28) |
| SN 2010jl | | | 0.9 | 20 | (86) |
| PTF 11qcj | 7.1 | 2.6 | 13 | $6.3 \times 10^{-2}$ | (89) |
| SN 1998bw | 0.45 | 100 | 15 | $3 \times 10^{-7}$ | (17, 20, 90) |
| SN 2007bg | 0.23 | 17 | 10 | $1.3 \times 10^{-3}$ | (17, 91) |
| SN 2006aj | $6.8 \times 10^{-2}$ | 200 | 6.2 | $< 9 \times 10^{-6}$ | (92, 93) |
| SN 1987A (ring) | $5.9 \times 10^{-6}$ | 17 | 70 | $5 \times 10^{-3}$ | (12, 94) |
| SN 1987A (inner) | | | 0.1 | $1 \times 10^{-3}$ | (94) |